\documentclass[epj]{svjour}
\usepackage{amsfonts}
\usepackage{amssymb}
\usepackage{amsmath}
\usepackage{multicol}
\usepackage{graphicx}
\usepackage{float}
\usepackage{caption}
\usepackage{graphics}
\usepackage{array}
\usepackage{comment}
\usepackage{wrapfig}
\usepackage{multirow}
\usepackage{tabularx}

\begin{document}

\title{Preheating and Reheating Constraints in Supersymmetric Braneworld
Inflation\thanks{%
Preheating and Reheating Constraints in Supersymmetric Braneworld Inflation}}
\subtitle{}
\author{K. El Bourakadi\inst{1}\thanks{\emph{Present address:}
k.elbourakadi@yahoo.com}, M. Bousder\inst{2}, Z. Sakhi\inst{1,2} \and M.
Bennai\inst{1,2} }
\institute{Physics and Quantum Technology Team, LPMC, Ben M'sik Faculty of Sciences,
Casablanca Hassan II University, Morocco \and LPHE-MS Laboratory department
of Physics, Faculty of Science, Mohammed V University in Rabat, Morocco}
\date{Received: date / Revised version: date}

%
\abstract{
We study the evolution of the Universe at early stages, we discuss also
preheating in the framework of hybrid braneworld inflation by setting
conditions on the coupling constants $\lambda $ and $g$\ for effective
production of $\chi$-particles. Considering the phase between the time
observable CMB scales crossed the horizon and the present time, we write 
reheating and preheating parameters $N_{re}$, $T_{re}$ and $N_{pre}$ in terms of the
scalar spectral index $n_{s}$, and prove that, unlike the reheating case, the
preheating duration does not depend on the values of the equation of state $\omega ^{\ast }$. We apply the slow-roll approximation in the high energy
limit to constrain the parameters of D-term hybrid potential. We show also that some inflationary parameters, in particular, the
spectral index $n_{s}$ demand that the potential parameter $\alpha$ is bounded as $\alpha \geq 1$ to be consistent with $Planck$'s data, while the ratio $r$ is in agreement with observation for $ \alpha \leq 1 $ considering high inflationary e-folds. We also propose an investigation of the brane tension effect on the reheating temperature. Comparing our results to recent CMB measurements, we study preheating and reheating
parameters $N_{re}$, $T_{re}$ and $N_{pre}$ in the Hybrid D-term inflation model in the range $0.8\leq \alpha\leq 1.1$\, and conclude that $T_{re}$ and $N_{re}$ require $\alpha \leq 1$, while for $N_{pre}$ the condition $\alpha \leq 0.9$ must be satisfied, to be compatible with $Planck$'s results.
\PACS{
      {PACS-key}{discribing text of that key}   \and
      {PACS-key}{discribing text of that key}
     } } 
\maketitle
\section{Introduction}

\label{intro}

After the brane-world inflation, preheating and reheating are considered to
be the stages at which elementary particles start populating the universe
that leads to matter creation at later times. In the brane-world
inflationary preheating scenario the four-dimensional Einstein equations
deviate from the standard cosmology \cite{G1}. By the proposal of a certain
type of compactification, the standard model particles are considered to be
on the three-dimensional brane, which makes gravitons propagate in the extra
dimension. This model indicates that we are living on the three-brane with a
positive tension embedded in five-dimensional anti-de Sitter bulk, which
appears in the Friedmann equation as an energy called brane tension. The
modifications due to additional dimension which contain the brane may have
implications on the reheating process. In the early stage of reheating
"preheating" \cite{G2}, the slow decreasing amplitude of inflaton field
coupled to $\chi$-field, results a strong amplification of massless $\chi$%
-particles through the non-perturbative decay of the inflaton \cite{G4}.%
\newline

While the single-field inflation models cannot reproduce the observational
data because they suffer from some fine-tuning problems on the parameters of
their potentials, such as the mass and the coupling constants \cite{G8}, the
hybrid inflationary model produces the observed temperature fluctuations in
the CMB, and show a great capacity to reproduce the experimental values of
all inflationary perturbation spectrum \cite{G9}. The difficulties of the
single-field inflation are also overcome in the context of hybrid
supersymmetric (SUSY) models \cite{G14}, since they agree with the observed
power spectrum of density perturbations very well \cite{G8}.\newline

In the context of supersymmetric theories, the D-term inflation looks more
promising since it avoids the problem associated with the inflaton mass \cite%
{G11}, and can be successfully implemented in the framework of supergravity
theories \cite{G12}. However, this type of inflation always ends with the
formation of cosmic strings \cite{G13}. The cosmic strings were analyzed in
the braneworld model, and it's been shown that Fayet--Iliopoulos (FI) term $%
\xi $ is responsible for this phenomenon \cite{Gb6}. As a consequence the
choice of the right value of (FI) term will be very important to avoid the
formation of the cosmic strings dominated in the conventional D-term
inflation models \cite{G12}.

After the supersymmetric inflation, a period that converts the stored energy
density in the inflaton to the thermal bath (a plasma of relativistic
particles) occurs \cite{G20}. This transition is known as reheating. During
this period, ordinary matter is produced as a result of inflaton field
energy loss. The scenario of reheating occurs when the inflaton oscillates
around the minimum of its potential and decay into new particles at the
radiation dominated period \cite{G22}. Along with this simple canonical
reheating model, a short phase include the non-perturbative processes \cite%
{G22} called preheating occurs. This scenario refers to the initial stage of
the reheating. Preheating is characterized by an\ exponential instability,
this instability corresponds to an exponential growth of occupation number
of quantum fluctuation, which is described by Mathieu equation. There are no
direct cosmological observables that are traceable to the phases that follow
inflation. However, one can consider the phase between the time observable
CMB scales crossed the horizon and the present time, as a way to achieve
information using indirect limits.\newline

Since we are interested in pre/reheating stages after SUSY brane inflation,
we are concerned to study the two phases durations quantified in terms of $e$%
-folds numbers $N_{pre}$ and $N_{re}$, in a certain interval of the equation
of state (EoS) $\omega ^{\ast }$. It is obvious that the parameter $\omega
^{\ast }$ have values larger than $-1/3$ , which is needed to end inflation 
\cite{G24}. This parameter is also assumed to be smaller than $+1$ to not
violate causality. On the other hand, its value at the beginning of the
radiation dominated era must be $+1/3$. Knowing that reheating start at $%
\omega ^{\ast }\geq 0$ \cite{G25}, therefore we will consider this parameter
with the choice of different values in that interval. We also calculate
preheating $e$-folds as a function of reheating temperature and analyze its
effects on preheating, and use the expressions for the effective D-term
potential to constrain pre/reheating parameters $\left(
T_{re},N_{re},N_{pre}\right) $, then compare the results with the
observations using the recently released $Planck-2018$ data \cite{G26}.
Knowing that the FI term was considered as a problem of\ D-term since the
CMB requires $\xi ^{\frac{1}{2}}\leq O(10^{15}-10^{16})\ GeV$ \cite{G27}%
.Below this range, the formation of cosmic strings becomes more important.
We conclude that\ the\ highest value possible of reheating temperature $%
T_{re}$ can favor a much higher $e$-folds number of preheating. Considering
the D-term hybrid model we study the pre/reheating durations, and take into
account values of FI term that doesn't lead to cosmic strings formation.%
\newline

Our paper is organized as follows: In the next section, we review the hybrid
braneworld inflation and reheating, and briefly discuss the preheating
mechanism in hybrid brane inflation. In section $3$ we derive the reheating
temperature and duration, we also derive the preheating duration in section $%
4$, finally, we constrain the parameters of D-term SUSY model and the
pre/reheating $e$-folds along with reheating temperature in section $5$. The
last section is devoted to a conclusion.

\section{Braneworld Hybrid inflation and preheating}

\label{sec:1}

\subsection{Formalisme of Braneworld inflation}

\label{sec:2} According to the braneworld model, the $(d+1)-$dimensional
anti-de Sitter space-time which we call the bulk, contains all the
observable universe embedded in 3-brane along with gravity and particle
fields. $RSII$ model is an inflation scenario of the braneworld scenario
with $d=4$. This means that $3-$brane is embedded in the five-dimensional $%
AdS$ space \cite{Ga1}, as a consequence the action of the field equation in $%
RSII$ model are written as \cite{Ga2}:

\begin{equation}
S_{RSII}=\int dx^{5}\sqrt{-g_{5}}\left[ \frac{M_{5}^{3}}{2}R_{b}+\Lambda _{5}%
\right] +\int dx^{4}\sqrt{-g_{4}}L_{brane},
\end{equation}%
where

\begin{equation}
L_{brane}=L_{matter}+\tilde{T}=\frac{1}{2}\left\vert \partial \Phi
\right\vert ^{2}-V+\tilde{T}.
\end{equation}%
$R_{b}$ is the bulk spacetime scalar curvature, with the bulk metric $g_{5}$%
, $M_{5}$ and $\Lambda _{5}$\ are the five-dimensional Planck mass and the
cosmological constant. The second term of $RSII$ action is the Lagrangian
density of ordinary matter on the brane, $V$ is the scalar potential, and $%
\tilde{T}$ is the brane tension. In this cosmological scenario,\textbf{\ }%
the metric projected onto the brane is a spatially flat
Friedmann--Robertson--Walker model, the Friedmann equation on the brane has
the generalized form \cite{Ga1b,Ga2}:

\begin{equation}
H^{2}=\frac{8\pi }{3M_{p}^{2}}\rho \left( 1+\frac{\rho }{2\tilde{T}}\right) .
\label{3}
\end{equation}

During inflation and preheating the term $\rho $ is expected to play
important role in the evolution of the Universe. In order to have the
inflationary dynamical equations leading to an accelerated expansion in the
high energy limit, we approximate $\rho\gg 2\tilde{T},$ this condition holds
until the end of preheating, considering the slow-roll conditions the energy
density can be approximated by $\rho \approx V$, and the Friedmann equation
finally takes the form :

\bigskip 
\begin{equation}
H^{2}=\frac{4\pi }{3M_{p}^{2}}\frac{V^{2}}{\tilde{T}}.
\end{equation}

The inflaton field confined on the brane satisfies the Klein-Gordon equation
given by

\begin{equation}
\ddot{\phi}+3H\dot{\phi}+V^{\prime }(\phi )=0,
\end{equation}%
with $V^{\prime }(\phi )=\partial V/\partial \phi ,$ we may also write the
spectral index $n_{s}$ and the ratio of tensor to scalar perturbations $r$
as functions of the slow-roll parameters $\epsilon _{k}$\ and $\eta _{k}$\ :

\begin{equation}
n_{s}=-6\epsilon +2\eta +1,~r\simeq 26\epsilon _{k}.
\end{equation}

The previous equations are very important tools of the braneworld RSII model
which have been used to solve some of the standard model problems.

\subsection{Hybrid inflation and reheating}

\label{sec:3} One reason to be interested in hybrid inflation is that it can
be implemented in supersymmetric theories \cite{Ga4}. The hybrid inflation
model which we study is defined by the potential \cite{Ga5}

\begin{equation}
V(\phi ,\sigma )=\frac{1}{4\lambda }\left( M^{2}-\lambda \sigma ^{2}\right)
^{2}+\frac{1}{2}m^{2}\phi ^{2}+\frac{1}{2}g^{2}\phi ^{2}\sigma ^{2}.
\end{equation}

The scalar fields $\phi $ and $\sigma $ have masses $m$ and $M$, the
potential has the symmetry $\sigma \leftrightarrow -\sigma $ at large values
of the field $\phi $. The potential has a maximum at $\phi =\sigma =0,$\
when the field $\phi $ value is small,\ and a global minimum at $\phi
=0,\sigma =\sigma _{0}=M/\sqrt{\lambda }$, where the symmetry is broken. The
equations of motion for the homogeneous fields are

\begin{eqnarray}
\ddot{\phi}+3H\dot{\phi} &=&-(m^{2}+g^{2}\sigma ^{2})\phi , \\
\ddot{\sigma}+3H\ddot{\sigma} &=&(M^{2}-g^{2}\sigma ^{2}-\lambda \sigma
^{2})\phi ,
\end{eqnarray}%
with $\dot{\phi}\equiv \partial \phi /\partial t$ and $\ddot{\phi}\equiv
\partial ^{2}\phi /\partial t^{2}$, at large $\phi $ the motion begins, the
effective mass of the $\sigma $ field $m_{\sigma }^{2}=g^{2}\phi ^{2}+M^{2}$
get also large. After the slow-roll of the field $\phi $, and just before
the end of\ inflation, $\phi $ acquires the critical value $\phi _{c}=M/g$,
and the field fluctuate at the minimum of the potential $\sigma =0$, then
the symmetry breaking phase transition caused by the massless $\sigma $
field fluctuations ends inflation. This transition could either end
instantaneously if the mass $M$ of the $\sigma $ field is larger than the
rate of expansion $H$, or if $M$ is of the order of $H,$ the transition will
be very slow which will make inflation have a few more $e$-folds after the
phase transition \cite{Ga7}.

At $\sigma =0$ the potential of inflaton field became $V(\phi
)=M^{4}/4\lambda +m^{2}\phi ^{2}/2$, considering the case where $m^{2}\ll
g^{2}M^{2}/\lambda $ the vacuum energy will dominate and the Hubble constant
at the time of the phase transition will be given by

\begin{equation}
H_{0}^{2}=\frac{\pi }{3\lambda ^{2}}\frac{M^{8}}{M_{p}^{2}}\frac{1}{\tilde{T}%
},
\end{equation}%
the slow-roll condition imposes the fact that we should neglect $\ddot{\phi}$
in the inflaton equation of motion, to finally\ have $3H\dot{\phi}\simeq
-V^{\prime }(\phi ).$ When the scalar field $\phi $ decreases below$\ \phi
_{c}=M/g\ $it prepares for the process of reheating.

\paragraph{Reheating Mechanism}

Reheating is an important phase that describes the production of standard
model particles, it occurs at the end of preheating with a decay rate $%
\Gamma =\Gamma \left( \phi \rightarrow \chi \chi \right) +\Gamma \left( \phi
\rightarrow \psi \psi \right) $, this means that the inflaton field $\phi $
can decay into bosons and fermions particles. Using Eq.(3) and the
expression of the energy density at the reheating epoch with $\rho =\rho
_{re}$\ \cite{G25}:

\begin{equation}
\rho _{re}=\frac{\pi ^{2}}{30}g_{re}T_{re}^{4},
\end{equation}%
one should find

\begin{equation}
H^{2}\simeq \frac{4\pi ^{5}}{3\times 30^{2}}\frac{g_{re}^{2}T_{re}^{8}}{%
M_{p}^{2}\tilde{T}}.
\end{equation}

We introduce the ratio between the Hubble parameter at phase transition and
the reheating phase given by

\begin{equation}
\left( \frac{H_{0}}{H}\right) ^{2}=\frac{30^{2}}{4\pi ^{4}\lambda ^{2}}\frac{%
M^{8}}{g_{re}^{2}T_{re}^{8}},
\end{equation}%
this ratio depends inversely on the reheating temperature, knowing that
according to \cite{Ga7c} the minimal possible value of this temperature is
bounded by $\left( T_{\min ,re}\geq 4Mev\right) $. 
\begin{figure}[H]
\centering
\includegraphics[width=15cm]{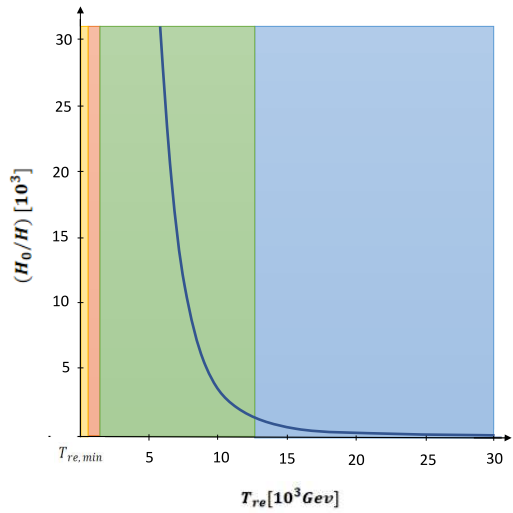}
\caption{Variation of the ratio $H_{0}/H$ as a function of reheating
temperature, the yellow region is ruled out by $BBN$(big bang
nucleosynthesis) and the orange region represents $100GeV$ of electroweak
scale, the light green region represents the inflation at phase transition
dominance, and light blue is the region of the reheating phase dominance.}
\label{fig:1}
\end{figure}
Fig.1 shows that the ratio $H_{0}/H$ decreases very rapidly when we consider
higher temperatures of reheating. In other words, we can say that for lower
values of reheating temperature where $H_{0}/H$ $>1,$ the difference between 
$H_{0}$ and $H$\ is considerably high, which means that the effect of
inflation at the phase transition is dominant in this region. At $%
H_{0}/H\leq 1$ we realize that reheating effects became much more important,
for that reason we suppose that this condition is accompanied by a new type
of expansion where the comoving Hubble scale $\left( aH\right) ^{-1}$ begin
to increase, which demand a certain number of $e$-foldings, that we will
prove to be a total duration written as $N_{pre}+f(\omega ^{\ast })N_{re}$\
in the remain sections.

In the next section, we will focus on the stage of preheating that we study
by introducing a $\chi $ field coupled to the potential $V(\phi ,\sigma )$.

\subsection{Preheating in hybrid inflation}

\label{sec:4} Particle production occurs when the fields oscillate around
their minimum. The behavior of the fields after the end of inflation could
describe the explosive preheating with a production of $\phi $ and $\sigma $
particles in hybrid inflation. However, we will study the particle
production behavior considering an extra scalar field $\chi $, coupled to
both of the previous fields \cite{Ga7d}

\begin{equation}
V(\chi )=\frac{1}{2}l_{1}^{2}\phi ^{2}\chi ^{2}+\frac{1}{2}l_{2}^{2}\sigma
^{2}\chi ^{2}.
\end{equation}

The explosive production of $\chi $ particles occurs for certain values of $%
l_{1}$ and $l_{2}$. The equations of motion of quantum fluctuations for\ $%
\chi $ field is given by

\begin{equation}
\ddot{\chi}_{k}+3H\dot{\chi}_{k}+(\frac{k^{2}}{a^{2}}+m_{\chi }^{2})\chi
_{k}=0,
\end{equation}%
the effective masse is written as

\begin{equation}
m_{\chi }^{2}=l_{1}^{2}\phi ^{2}+l_{2}^{2}\sigma ^{2}.
\end{equation}

The rate of expansion of the universe plays an important role in ending the
parametric resonance regime, particle production will be ended by the
redshifted modes with momentum $k/a$ that will fall out of the resonance
band because of the expansion. According to the previous section, there are
two fundamental frequencies near the minimum of the potential: $\nu _{\sigma
}=\sqrt{2}M$ and $\nu _{\phi }=gM/\sqrt{\lambda }$, that corresponds to the
two critical values of the fields $\sigma $ and $\phi $: $\sigma _{0}=M/%
\sqrt{\lambda }$ and $\ \phi _{c}=M/g$. Particle production in $\chi $ field
that interacts with the fields $\phi $ and $\sigma $ given by Eq.(15)
require an amplified fluctuations of this field around $\phi =0$ while $%
\sigma \simeq \sigma _{0}$, this demand that the induced mass from the
symmetry breaking field $\sigma $ must be much smaller than the
corresponding oscillations from the $\phi $ field $l_{1}^{2}\phi ^{2}\gg
l_{2}^{2}\sigma _{0}^{2}$, which at the beginning of preheating corresponds
to

\begin{equation}
\lambda l_{1}^{2}\gg g^{2}l_{2}^{2}.
\end{equation}%
Using the solution of the Klein-Gordon equation\ given by $\phi (t)\approx
\Phi (t)sin(\nu _{\phi }t),$ we can determine that the explosive production
of $\chi -$particles will end when the amplitude of oscillations of the $%
\phi $ field becomes $\Phi (t)<\left( l_{2}g\right) /l_{1}\sqrt{\lambda }$.

The Mathieu equation which corresponds to the preheating process is given by

\begin{equation}
\chi _{k}^{\prime \prime }+(A(k)-2qcos(2z))\chi _{k}=0,
\end{equation}
with "$^{\prime \prime }\equiv \partial ^{2}/\partial z^{2}$" is the
derivative with respect to the parameter $z,$ and

\begin{eqnarray}
A_{k} &=&\frac{k^{2}}{a\nu _{\phi }^{2}}+2q, \\
q &=&\frac{l_{1}^{2}M^{2}}{g^{2}\nu _{\phi }^{2}}\frac{\Phi ^{2}}{4}.
\end{eqnarray}

Taking into consideration the previous condition Eq.(17), at the initial
step of preheating, Eq.(20) must satisfy $q_{0}=\left( \lambda
l_{1}^{2}/g^{4}\right) \Phi ^{2}/4,$ for $\lambda \gg g^{2}$ we can find
natural values of the parameter $l_{1}$ that ensure $q_{0}\gg 1$. This will
allow an explosive production of $\chi -$particles, which mean we are in the
broad parametric resonance region, that corresponds to an exponential growth
of quantum fluctuations occupation numbers, given by \cite{G4}

\begin{equation}
n_{k}(t)\simeq exp(2\mu _{k}z),
\end{equation}%
with a frequency $\omega _{k}$

\begin{equation}
\omega _{k}^{2}=\frac{k^{2}}{a^{2}}+l_{1}^{2}\phi ^{2}.
\end{equation}

The choice of natural values of couplings will cause an efficient particle
production process, because of a large value of the growth parameter $\mu
_{k}$. Furthermore, for a large range of couplings we will enter the region
of stochastic resonance \cite{G4}.

\section{Reheating duration and \textbf{temperature}}

\label{sec:5} The reheating duration can be extracted considering the phase
between the time observable CMB scales crossed the horizon and the present
time. Deferents eras occurred throughout this length of time that can be
described by the two following equations \cite{G25}:

\begin{equation}
\frac{k}{a_{0}H_{0}}=\frac{a_{k}}{a_{end}}\frac{a_{end}}{a_{re}}\frac{a_{re}%
}{a_{eq}}\frac{a_{eq}H_{eq}}{a_{0}H_{0}}\frac{H_{k}}{H_{eq}},  \label{23}
\end{equation}%
the pivot scale for a specific experiment is parametrized by $k$\ \cite{G25},%
$\ N_{k}$ is the $e$-folds of the inflation era, $N_{pre},N_{re}$ and $%
N_{RD} $ respectively correspond to the reheating and radiation-domination
era durations. Reheating is characterized by the temperature $T_{re}$ and a
number of $e$-folds $N_{re}=\ln \left( a_{re}/a_{end}\right) $ occurring
before the radiation dominated era. Using $\rho \propto a^{-3(1+\omega
^{\ast })}$, the reheating epoch is described by :

\begin{equation}
\frac{\rho _{end}}{\rho _{re}}=\left( \frac{a_{end}}{a_{re}}\right)
^{-3(1+\omega ^{\ast })},  \label{24}
\end{equation}%
where $\rho _{end}$ and $a_{end}$ corresponds to the end of inflation and $%
\rho _{re}$ and $a_{re}$ corresponds to the end of reheating. As a result :

\begin{equation}
N_{re}=\frac{1}{3\left( 1+\omega ^{\ast }\right) }\ln \left( \frac{\rho
_{end}}{\rho _{re}}\right) ,
\end{equation}%
knowing that $\rho _{end}=(6/5)V_{end}$\ in the Braneworld case \cite{G25a}\
and $\rho _{re}=\left( \pi ^{2}/30\right) g_{re}T_{re}^{4}$, we replace
every energy density by its expression to obtain :

\begin{equation}
N_{re}=\frac{1}{3\left( 1+\omega ^{\ast }\right) }\ln \left( \frac{6}{5}%
\frac{30V_{end}}{\pi ^{2}g_{re}T_{re}^{4}}\right) .  \label{26}
\end{equation}%
The reheating temperature and the actual temperature are related as \cite%
{G24} :

\begin{equation}
T_{re}=T_{0}\left( \frac{a_{0}}{a_{eq}}\right) e^{N_{RD}}\left( \frac{43}{%
11g_{re}}\right) ^{\frac{1}{3}},  \label{27}
\end{equation}%
using the following expression in the previous equation :

\begin{equation}
\frac{a_{0}}{a_{eq}}=\frac{a_{0}H_{k}}{k}e^{-N_{k}}e^{-N_{re}}e^{-N_{RD}},
\label{28}
\end{equation}%
gives the result

\begin{equation}
N_{re}=\frac{4}{1-3\omega ^{\ast }}\left[ -\frac{1}{4}\ln \left( \frac{36}{%
\pi ^{2}g_{re}}\right) -\frac{1}{3}\ln \left( \frac{11g_{re}}{43}\right)
-\ln \left( \frac{k}{a_{0}T_{0}}\right) -\ln \left( \frac{V_{end}^{\frac{1}{4%
}}}{H_{k}}\right) -N_{k}\right] ,
\end{equation}%
it is important to notice that the reheating duration is not defined in the
value of the\ equation of state $\omega ^{\ast }=1/3$. We can simplify the
previous expression considering $g_{re}\approx 200$\ for the Braneworld case 
\cite{G25a}\ and the pivot scale $0.05Mpc$\ \cite{G25}, and other numerical
values From \cite{G26} : $M_{pl}=2.435\times 10^{18}GeV$, $a_{0}=1$, $%
T_{0}=2.725K$ :

\begin{equation}
N_{re}=\frac{4}{1-3\omega ^{\ast }}\left[ 6.61-\ln \left( \frac{V_{end}^{%
\frac{1}{4}}}{H_{k}}\right) -N_{k}\right] .
\end{equation}%
The estimation for $N_{re}$ is not accessible for $\omega ^{\ast }=1/3$.
Note that inconsistency in $N_{re}$ exists for this case because we define
the beginning of radiation dominance when $\omega ^{\ast }$ reaches $1/3$.

From Eqs. (\ref{27}), (\ref{28}) we obtain

\begin{equation}
T_{re}=\left( \frac{43}{11g_{re}}\right) ^{\frac{1}{3}}\left( \frac{%
a_{0}T_{0}}{k}\right) H_{k}~e^{-N_{k}}e^{-N_{re}},
\end{equation}%
using the previous equation with Eq.(\ref{26}), we can find the reheating
temperature as

\begin{equation}
T_{re}=\left[ \left( \frac{43}{11g_{re}}\right) ^{\frac{1}{3}}\frac{%
a_{0}T_{0}}{k}H_{k}~e^{-N_{k}}\left( \frac{36V_{end}}{\pi ^{2}g_{re}}\right)
^{\frac{-1}{3(1+\omega ^{\ast })}}\right] ^{\frac{3(1+\omega ^{\ast })}{%
3\omega ^{\ast }-1}}.
\end{equation}%
To find $N_{re}$\ and $T_{re}$\ for a particular model, one needs to compute 
$N_{k}$, $H_{k}$\ and $V_{end}$.

\section{Preheating duration}

\label{sec:6}

We start our computation for the preheating case, by introducing the ratio $%
a_{pre}/a_{pre}$ to Eq. (\ref{23}) that corresponds to preheating scale
factor. as result, we obtain an equation written in terms of $e$-folds

\begin{equation}
\ln \frac{k}{a_{0}H_{0}}=-N_{k}-N_{pre}-N_{re}+\ln \frac{a_{re}}{a_{0}}+\ln 
\frac{H_{k}}{H_{0}},
\end{equation}%
where $N_{pre}\equiv \ln \left( a_{pre}/a_{end}\right) $ is the number of $e$%
-folds between the end of inflation and the end of preheating. Assuming that
no entropy production took place after the completion of reheating, one can
write \cite{G24,G24b} 
\begin{equation}
\frac{a_{re}}{a_{0}}=\frac{T_{0}}{T_{re}}\left( \frac{43}{11g_{re}}\right) ^{%
\frac{1}{3}},
\end{equation}%
where $T_{0}$ is the current temperature of the Universe. Using Eq. (\ref{24}%
) and the expression of reheating density energy $\rho _{re}=(\pi
^{2}/30)g_{re}T_{re}^{4}$, we can determine an expression given by:

\begin{eqnarray}
N_{tot} &=&\dfrac{1}{f(\omega ^{\ast })}N_{pre}+N_{re},  \label{Nt} \\
&=&\dfrac{1}{f(\omega ^{\ast })}\left[ -\ln \left( \frac{k}{a_{0}T_{0}}%
\right) -\frac{1}{3}\ln \left( \frac{11g_{re}}{43}\right) -\frac{1}{4}\ln
\left( \frac{36}{\pi ^{2}g_{re}}\right) -\frac{1}{4}\ln \left( \frac{V_{end}%
}{H_{k}^{4}}\right) -N_{k}\right] .  \notag
\end{eqnarray}%
Eq. (\ref{Nt}) can be reduced to

\begin{equation}
N_{pre}+f(\omega ^{\ast })N_{re}=\left[ 6.61-\ln \left( \frac{V_{end}^{\frac{%
1}{4}}}{H_{k}}\right) -N_{k}\right] ,
\end{equation}%
here $f(\omega ^{\ast })=\left( 1-3\omega ^{\ast }\right) /4$, using Eq. 26,
the final form of $N_{pre}$ will be given as : 
\begin{equation}
N_{pre}=\left[ 6.61-\ln \left( \frac{V_{end}^{\frac{1}{4}}}{H_{k}}\right)
-N_{k}\right] -\frac{1}{12}\frac{\left( 1-3\omega ^{\ast }\right) }{\left(
1+\omega ^{\ast }\right) }\ln \left[ \frac{36\cdot V_{end}}{\pi
^{2}g_{re}T_{re}^{4}}\right] .  \label{Npre}
\end{equation}

According to \cite{Gb13,Gb13a}, if reheating is instantaneous the
temperature would be in the order of $10^{14}GeV$, taking into account this
maximum reheating temperature, this condition favors a much higher $e$-folds
number of preheating. Next we need to find the expressions of $N_{k},H_{k}$
and $V_{end}$, the tree parameters can be related to the scalar spectral
index $n_{s}$, and they can be connected directly to a model of inflation,
which means that we can use D-term model to study pre/reheating constraints
in SUSY brane-world inflation.

\section{Pre/Reheating constraints in supersymmetric braneworld inflation}

\label{sec:7} Preheating and reheating are now parametrized by $e$-folds
numbers $N_{pre}$ and $N_{pre}$, but we do not expect that both durations
perform the same behavior. From Eqs. (30,36), we observe that pre/reheating
are expressed as functions of tree parameters and a constant equation of
state (EoS) $\omega ^{\ast }$. For inflation to be ended the value of $%
\omega ^{\ast }$ should be larger than $-1/3$, in order to satisfy the
condition of density energy dominance and preserve the causality, $\omega
^{\ast }$ must be smaller than $1$. During reheating the (EoS) increased
from $0$ to $1/3,$ this will attract attention to study pre/reheating
durations as functions of (EoS) $\omega ^{\ast }$, to test the effects of
(EoS) parameter on both phases. Since $N_{k},H_{k}$ and $V_{end}$ can be
related to the scalar spectral index $n_{s}$, we can Therefore use the
observational data to place constraints on the pre/reheating durations for
the supersymmetric D-term brane inflation.

\subsection{D-term hybrid inflation}

\label{sec:8}

D-term inflation was first proposed to solve the fine-tuning problem \cite%
{Gb7}, the D-term require to introduce a U(1) symmetry with a
Fayet--Iliopoulos (FI) term $\xi $, in order to have a broken supersymmetric
D-term inflation \cite{Gb8}. The D-term Hybrid inflation was derived to the
form \cite{Gb12}:

\begin{equation}
V=\frac{g^{2}\xi ^{2}}{2}\left( 1+\alpha ~\ln \left( \frac{\lambda \phi ^{2}%
}{\Lambda ^{2}}\right) \right) ,  \label{V}
\end{equation}%
where $\alpha =g^{2}/$\ $16\pi ^{2}$\ and $\Lambda $\ is a renormalization
mass scale, considering that just before the end of inflation at $\phi
_{end},$\ inflation reached a critical value $\phi =\phi _{c},$\ following
the results of \cite{G9}, $\phi _{c}$\ is given as

\begin{equation*}
\phi _{c}=\frac{g}{\lambda }\sqrt{\xi },
\end{equation*}%
On the other hand, in the high-energy limit $\left( V\gg 2\tilde{T}\right) $%
\ the well-known slow-roll parameters \cite{Ga8b} \bigskip are given by $%
\epsilon _{k}=\left( M_{p}^{2}\tilde{T}V^{\prime 2}\right) /\left( 4\pi
V^{3}\right) $\ and $\eta _{k}=\left( M_{p}^{2}\tilde{T}V^{\prime \prime
}\right) /\left( 4\pi V^{2}\right) ,$\ where $\epsilon _{k}\ll 1,\eta
_{k}\ll 1$. Assuming that $g^{2}\xi ^{2}/4\tilde{T}\gg 1$\ in Eq.(\ref{V}),\
therefore $g^{2}\xi ^{2}$\ dominates and $V^{\prime }/V=\alpha /\phi _{k}$.\
We calculate $\epsilon _{k}$\ and $\eta _{k}$ as

\begin{eqnarray}
\mathbf{\ }\epsilon _{k} &=&\frac{\alpha ^{2}}{4\pi }\frac{\tilde{T}}{%
g^{2}\xi ^{2}/2}\frac{M_{p}^{2}}{\phi _{k}^{2}},  \label{e1} \\
\eta _{k} &=&\frac{\alpha }{4\pi }\frac{\tilde{T}}{g^{2}\xi ^{2}/2}\frac{%
M_{p}^{2}}{\phi _{k}^{2}},  \label{e2}
\end{eqnarray}%
in order to constrain the parameters of the potential in Eq.(\ref{V}),\ the
expression of the scalar field at the end of inflation\ is obtained by
solving the equation $\left\vert \eta \left( \phi _{end}\right) \right\vert
=1$\ knowing that $\eta \gg \epsilon ,$

\begin{equation}
\phi _{end}=\frac{M_{p}\sqrt{\tilde{T}}}{4\pi ^{\frac{3}{2}}\xi }.
\label{end}
\end{equation}%
Considering the condition $\phi _{c}\geq $\ $\phi _{end},$\ we obtain

\begin{equation}
\tilde{T}\leq \frac{16\pi ^{3}g^{2}\xi ^{3}}{M_{p}^{2}~\lambda ^{2}},
\label{T}
\end{equation}%
based on Eq.(\ref{T}) and the analysis in \cite{G10}, it can be found that 

\begin{eqnarray}
\lambda ^{2} &<&8\pi ^{2}\sqrt{3P_{s}(k)}, \\
\xi  &<&\frac{\sqrt{3P_{s}(k)}}{8\pi }M_{p}^{2}, \\
\tilde{T} &<&\frac{g^{2}\left( 3P_{s}(k)\right) ^{3/2}}{32\lambda ^{2}}%
M_{p}^{4},
\end{eqnarray}%
assuming that $\alpha =g^{2}/$\ $16\pi ^{2}$\ is bounded by $0.8\leq \alpha
\leq 1.1$, and the power spectrum of the curvature perturbations is measured
as $P_{s}\simeq A_{s}=2.196_{-0.06}^{+0.05}\times 10^{-9}$ from the recent
Planck results \cite{G26}, numerically we found 

\begin{eqnarray}
\lambda ^{2} &\leq &0.08, \\
\xi &\leq &1.9\times 10^{31}, \\
\tilde{T} &\leq &\frac{7.02\times 10^{61}}{\lambda ^{2}}GeV^{4}.
\end{eqnarray}

\bigskip The number of $e$-foldings $N_{k}$\ during inflation from the time
when mode $k$\ leaves the horizon to the end of inflation can be obtained in
the high-energy limit as

\begin{equation}
N_{k}\simeq \frac{4\pi }{M_{p}^{2}\tilde{T}}\int_{\phi _{end}}^{\phi _{k}}%
\frac{V^{2}}{V^{\prime }}d\phi ,  \label{N}
\end{equation}%
which we can apply to the potential Eq.(\ref{V}) considering that $g^{2}\xi
^{2}$\ dominates and $V^{2}/V^{\prime }=g^{2}\xi ^{2}\phi _{k}/\alpha ,$

\begin{equation}
N_{k}=\frac{4\pi }{\alpha \tilde{T}M_{p}^{2}}\frac{g^{2}\xi ^{2}}{2}\left(
\phi _{k}^{2}-\phi _{end}^{2}\right) ,
\end{equation}%
considering the approximation $\phi _{k}\gg \phi _{end}$\ , we obtain :

\begin{equation}
N_{k}=\frac{4\pi }{\alpha \tilde{T}}\frac{g^{2}\xi ^{2}}{2}\frac{\phi
_{k}^{2}}{M_{p}^{2}},
\end{equation}%
from the previous equation, we can derive an expression for the inflaton
field at the horizon crossing $\phi _{k}$

\begin{equation}
\phi _{k}^{2}=\frac{\alpha \tilde{T}M_{p}^{2}}{2\pi g^{2}\xi ^{2}}N_{k}~.
\end{equation}%
The scalar spectral index defined as \cite{Ga8c} $n_{s}-1=-6\epsilon
_{k}+2\eta _{k}$\ can be used to find 
\begin{equation}
\frac{M_{p}^{2}}{\phi _{k}^{2}}=\frac{2\pi g^{2}\xi ^{2}/2}{\tilde{T}}\frac{%
\left( 1-n_{s}\right) }{\alpha \left( 3\alpha -1\right) },  \label{cle}
\end{equation}%
after replacing Eq.(\ref{cle}) in Eqs. (\ref{e1}), (\ref{e2}) and (\ref{N}),
one should get

\begin{eqnarray}
\epsilon _{k} &=&\frac{\alpha \left( 1-n_{s}\right) }{2\left( 3\alpha
-1\right) }, \\
\eta _{k} &=&\frac{\left( 1-n_{s}\right) }{2\left( 3\alpha -1\right) }, \\
N_{k} &=&\frac{2\left( 3\alpha -1\right) }{1-n_{s}}.
\end{eqnarray}

Since the parameter $r$\ can be written as $r\simeq 24\epsilon _{k}$, we can
explicit the dependence of parameter $\alpha $\ on inflationary observables,
to perform this analysis we use 

\begin{eqnarray}
n_{s}-1 &=&\frac{1-3\alpha }{N_{k}}, \\
\mathbf{\ }r &=&\frac{12\alpha }{N_{k}},
\end{eqnarray}%
considering inflation $e$-folds in the range $50\leq N_{k}\leq 80,$\ we
selected the following values $\tilde{T}\simeq 10^{58}GeV^{4},~\xi \simeq
10^{30},~0.8\leq \alpha \leq 1.1~and~\lambda \simeq 0.08$\ and plotted $r$\
as a function $n_{s}$\ 
\begin{table}[H]
\centering%
\begin{tabular}{||ccc||}
\hline
Inflationary $e$-folds $N_{k} $ ($\alpha =0.8$) & $n_{s}$ & $r$ \\%
[0.5ex] \hline\hline
50 & 0.972 & 0.19 \\ 
60 & 0.976 & 0.16 \\ 
70 & 0.980 & 0.13 \\ 
80 & 0.982 & 0.12 \\[1ex] \hline
\end{tabular}%
\caption{Braneworld-inflationary parameters $n_{s}$\ and $r$\ for $%
50<N_{k}\leq 80$\ in the case of $\protect\alpha =0.8$ }
\label{table:1}
\end{table}
\begin{table}[H]
\centering%
\begin{tabular}{||ccc||}
\hline
Inflationary $e$-folds $N_{k} $ ($\alpha =0.9$) & $n_{s}$ & $r$ \\%
[0.5ex] \hline\hline
50 & 0.966 & 0.21 \\ 
60 & 0.971 & 0.18 \\ 
70 & 0.975 & 0.15 \\ 
80 & 0.978 & 0.13 \\[1ex] \hline
\end{tabular}%
\caption{Braneworld-inflationary parameters $n_{s}$\ and $r$\ for $%
50<N_{k}\leq 80$\ in the case of $\protect\alpha =0.9$ }
\label{table:2}
\end{table}
\begin{table}[H]
\centering%
\begin{tabular}{||ccc||}
\hline
Inflationary $e$-folds $N_{k} $ ($\alpha =1$) & $n_{s}$ & $r$ \\%
[0.5ex] \hline\hline
50 & 0.960 & 0.24 \\ 
60 & 0.966 & 0.20 \\ 
70 & 0.971 & 0.17 \\ 
80 & 0.975 & 0.15 \\[1ex] \hline
\end{tabular}%
\caption{Braneworld-inflationary parameters $n_{s}$\ and $r$\ for $%
50<N_{k}\leq 80$\ in the case of $\protect\alpha =1$ }
\label{table:3}
\end{table}
\begin{table}[H]
\centering%
\begin{tabular}{||ccc||}
\hline
Inflationary $e$-folds $N_{k} $ ($\alpha =1.1$) & $n_{s}$ & $r$ \\%
[0.5ex] \hline\hline
50 & 0.954 & 0.26 \\ 
60 & 0.961 & 0.22 \\ 
70 & 0.967 & 0.18 \\ 
80 & 0.971 & 0.16 \\[1ex] \hline
\end{tabular}%
\caption{Braneworld-inflationary parameters $n_{s}$\ and $r$\ for $%
50<N_{k}\leq 80$\ in the case of $\protect\alpha =1.1$}
\label{table:4}
\end{table}

\begin{figure}[H]
\centering
\includegraphics[width=15cm]{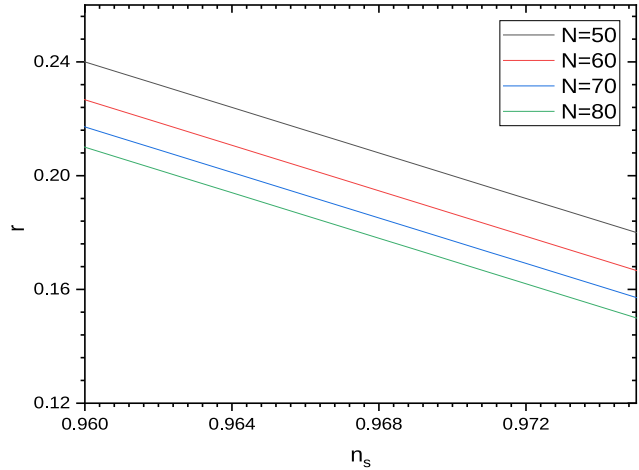}
\caption{ $r$ vs. $n_{s}$ for D-term braneworld inflation.}
\label{fig:2}
\end{figure}

\textbf{\ }Figure \ref{fig:2} shows that the parameter $r$\ is a decreasing
function with respect to $n_{s}$. We observe that $\alpha $\ must be$\ 1\leq
\alpha \leq 1.1$\ in order to have inflation $e$-folds with scalar spectral
index $n_{s}$\ compatible with Planck's results$.$\ We note also that, the
tensor to scalar ratio is in agreement with the observations for $\alpha
\leq 1$, considering higher inflation $e$-folds number $N_{k}\geq 70.$

The quantum fluctuations of the scalar field lead also to fluctuations in
the metric. In this way, one can define the amplitude of tensor
perturbations as \cite{Ga8}\ 
\begin{equation}
P_{h}(k)=\frac{64\pi }{M_{p}^{2}}\left( \frac{H}{2\pi }\right) ^{2}F^{2}(x),
\end{equation}%
where $x=HM_{p}\sqrt{\frac{3}{4\pi \tilde{T}}}$ and $F^{2}(x)=\left( \sqrt{%
1+x^{2}}-x^{2}\sinh ^{-1}\left( \frac{1}{x}\right) \right) ^{-1}.$ Note that
in the high-energy limit $\left( V\gg 2\tilde{T}\right) ,\ F^{2}(x)\approx 
\frac{3}{4}x=\frac{3}{4}\frac{V}{\tilde{T}}$. Thus allowing

\begin{equation}
P_{h}(k)=\frac{24}{\pi M_{p}}H^{3}\sqrt{\frac{3}{4\pi \tilde{T}}}.
\label{41}
\end{equation}

\bigskip The ratio of tensor to scalar perturbations $r$ is given by

\begin{equation}
r=\frac{P_{h}(k)}{P_{s}(k)},  \label{42}
\end{equation}%
the inflation parameter $r$\ can be written as $r\simeq 24\epsilon _{k}$. By
using Eqs. (\ref{41},\ref{42}), the expression of $H_{k}$\ is given by

\begin{equation}
H_{k}=\left( \pi M_{p}A_{s}\epsilon _{k}\sqrt{\frac{4\pi \tilde{T}}{3}}%
\right) ^{\frac{1}{3}},
\end{equation}%
at the pivot scale $k=0.05Mpc^{-1}$, the power spectrum amplitude measured
by Planck is $P_{\xi }\simeq A_{s}$, with $A_{s}=2.196_{-0.06}^{+0.05}\times
10^{-9}$ \cite{G26}.\ The brane tension is considered of the order of $%
10^{58}GeV$ \ \cite{Gb13}. Finally, for this model, we derive the following
expression 
\begin{equation}
H_{k}=\left( \pi M_{p}A_{s}\frac{\alpha \left( 1-n_{s}\right) }{2\left(
3\alpha -1\right) }\sqrt{\frac{4\pi \tilde{T}}{3}}\right) ^{\frac{1}{3}},
\end{equation}%
\ we can determine $V_{end}=V\left( \phi _{end}\right) $\ using the
expression of the scalar field at the end of inflation given by Eq.(\ref{end}%
).

We previously constrain the value of brane tension, in that sense, we will
study the effect of the brane tension $\tilde{T}$\ on reheating temperature$%
\ T_{re}$. In in the high energy limit $\rho \gg 2\tilde{T},$\ Eq.(\ref{3})
can be written as

\begin{equation}
H^{2}=\frac{4\pi }{3M_{p}^{2}}\frac{\rho ^{2}}{\tilde{T}},
\end{equation}%
considering $\rho _{re}=\left( \pi ^{2}/30\right) g_{re}T_{re}^{4}$\ we
obtain the following equation

\begin{equation}
T_{re}\ =\left( \frac{3\times 30^{2}}{4\pi ^{5}}\frac{M_{p}^{2}~H^{2}}{%
g_{re}^{2}}\tilde{T}\right) ^{\frac{1}{8}},  \label{a}
\end{equation}%
the value of the scalar field at the inflation $\phi _{end}$\ leads to an
expression of the Friedmann equation as :

\begin{equation}
H^{2}(end)=\frac{4\pi }{3M_{p}^{2}}\frac{V_{end}^{2}}{\tilde{T}},  \label{b}
\end{equation}%
we can computed the reheaing tempreture $T_{re}$\ as a function of the bran
tension $\tilde{T}$\ considering Eq.(\ref{a}) and (\ref{b}), to finally
obtain

\begin{equation}
T_{re}\ =\left( \frac{30^{2}}{\pi ^{4}}\frac{V_{end}^{2}}{g_{re}^{2}}\right)
^{\frac{1}{8}},
\end{equation}%
\begin{figure}[]
\centering
\includegraphics[width=15cm]{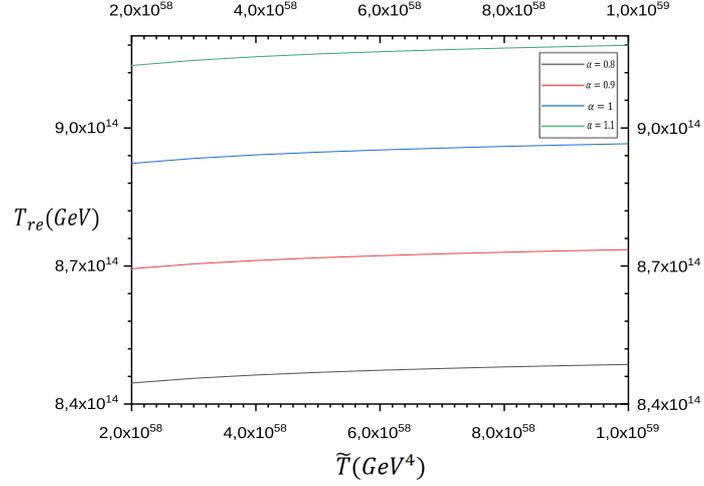}
\caption{ $T_{re}$ versus $\tilde{T}$ for different values of $\protect%
\alpha ,$ the black-line corresponds to $\protect\alpha =0.8,$ the red-line
corresponds to $\protect\alpha =0.9,$\ the blue-line corresponds to $\protect%
\alpha =1,$ and the green-line corresponds to $\protect\alpha =1.1.$\ }
\label{fig:3}
\end{figure}

Fig. (\ref{fig:3}) shows the evolution of the reheating temperature $T_{re}$%
\ as a function of brane tension $\tilde{T}$\ for different values of the
parameter $\alpha .$\ We remark that $T_{re}$\ has an increasing behavior as
we increase the tension $\tilde{T}$. We vary the tension $\tilde{T}$\ in the
range $\left( 2-10\right) \times 10^{58}GeV^{4},$\ for the case $\alpha =0.8$%
\ we get $T_{re}\sim \left( 8.44-8.48\right) \times 10^{14}GeV,$\ but when
we consider the case $\alpha =0.9$\ the temperature evolve around $%
T_{re}\sim \left( 8.69-8.73\right) \times 10^{14}GeV,$\ for $\alpha =1$\ we
obtain $T_{re}\sim \left( 8.92-8.96\right) \times 10^{14}GeV,$\ while the
case $\alpha =1.1$\ gives $T_{re}\sim \left( 9.13-9.17\right) \times
10^{14}GeV.$ \ 

In the following, we apply this formalism to study the variation of the
reheating temperature and pre/reheating $e$-folding number as a function of
the perturbation spectrum.

\subsection{Reheating case}

\label{sec:9}

The reheating phase is parametrized in terms of a duration $N_{re}$,
thermalization temperature $T_{re}$ at the equilibrium stat of reheating,
and equation of state $\omega ^{\ast }$. Note that the time evolution of the 
$\omega ^{\ast }(t)$ for different couplings was studied in Ref. \cite{G23}
and it was shown that it varies slightly for very short times during
different phases following inflation. Here, we suppose that $-1/3\leq \omega
^{\ast }\leq 1/3$.

\begin{center}
\begin{figure}[H]
\centering
\includegraphics[width=15cm]{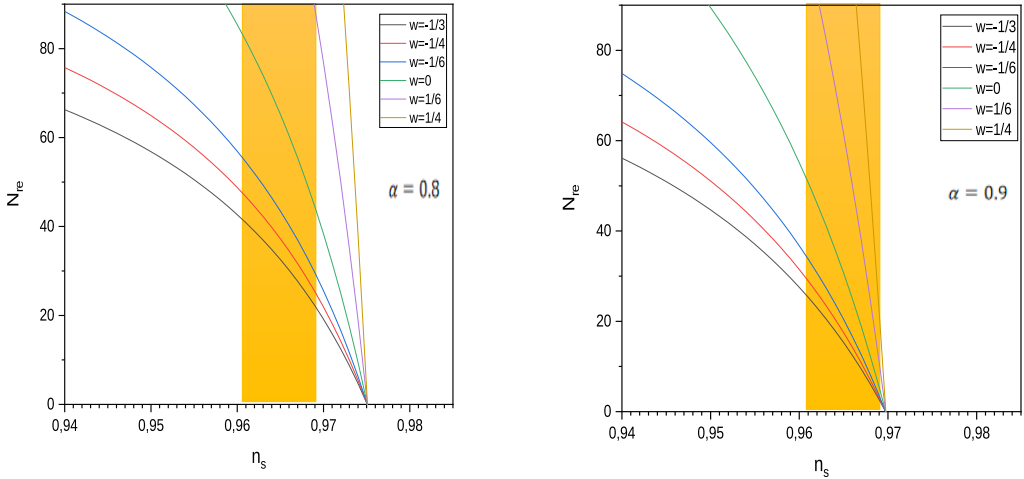}
\caption{Variation of $N_{re}$ as function of $n_{s}$ for $\protect\alpha %
=0.8$ and $\protect\alpha =0.9$ using differentes values of $\protect\omega %
^{\ast }$.}
\label{fig:4}
\end{figure}
\begin{figure}[H]
\centering
\includegraphics[width=15cm]{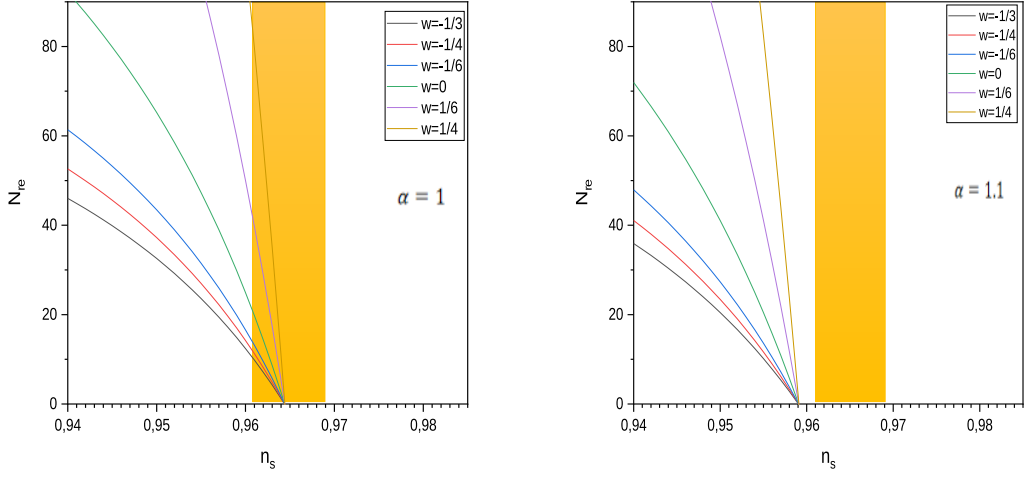}
\caption{Variation of $N_{re}$ as function of $n_{s}$ for $\protect\alpha =1$
and $\protect\alpha =1.1$ using differentes values of $\protect\omega ^{\ast
}$.}
\label{fig:5}
\end{figure}
\end{center}

In Figs. \ref{fig:4} and \ref{fig:5}, we have plotted the variation of
reheating $e$-folding number $N_{re}$ as a function of $n_{s}$ for different
values of $\omega ^{\ast }$, in the case of the D-term hybrid potential in
braneworld inflation. The vertical yellow region represents $Planck-2018$
bounds on $n_{s}=0.9649\pm 0.0042$ \cite{G26}. The point where all the lines
converge is called instantaneous reheating and is defined in the limit $%
N_{re}\rightarrow 0$. We observe that for $\alpha =1$ all the lines are
shifted toward the central value of $n_{s}$. The case $\alpha =1.1$ is
difficult to reconcile for all $\omega ^{\ast }$ in the bounds on $n_{s}$. 
\begin{figure}[H]
\centering
\includegraphics[width=15cm]{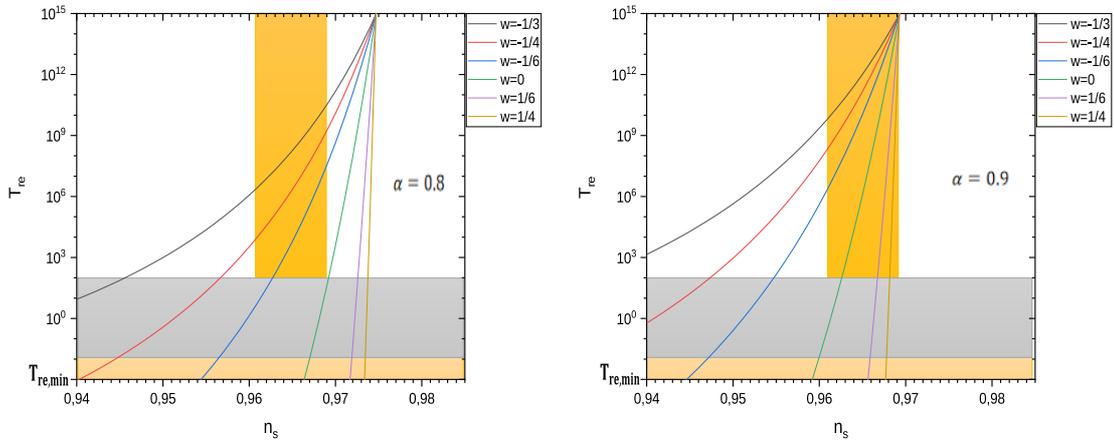}
\caption{Variation of $T_{re}$ as function of $n_{s}$ for $\protect\alpha %
=0.8$ and $\protect\alpha =0.9$ using differentes values of $\protect\omega %
^{\ast }$.}
\label{fig:6}
\end{figure}
\begin{figure}[H]
\centering
\includegraphics[width=15cm]{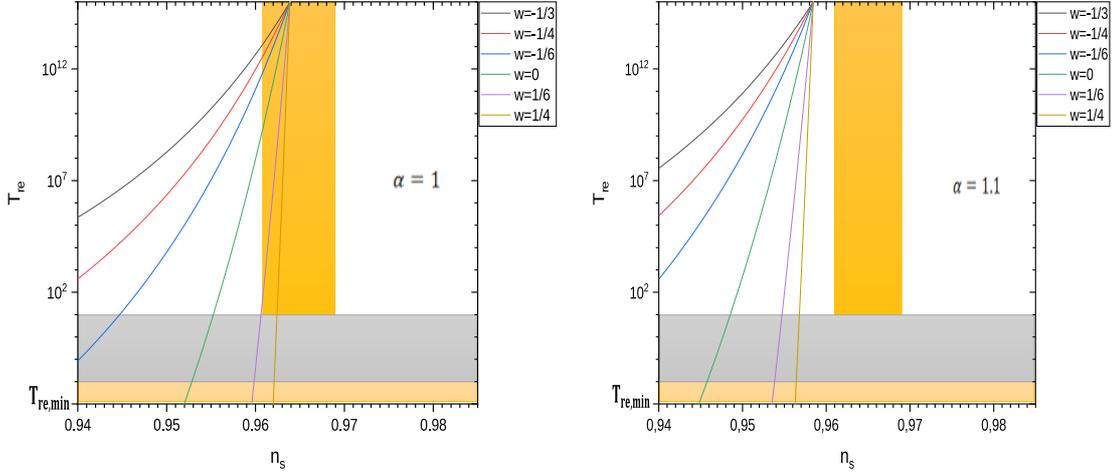}
\caption{Variation of $T_{re}$ as function of $n_{s}$ for $\protect\alpha =1$
and $\protect\alpha =1.1$ using differentes values of $\protect\omega ^{\ast
}$.}
\label{fig:7}
\end{figure}

In Figs. \ref{fig:6} and \ref{fig:7}, we should note that in all cases, the $%
T_{re}$\ converges around $10^{15}GeV$\ as may be required by GUT scale
baryogenesis models. Temperatures below the light gray region are ruled out
by BBN and light gray region represents $100~GeV$\ of electroweak scale. The
case $\alpha =1.1$\ is difficult to reconcile for $\omega ^{\ast }\leq 1/4$\
in the bounds on $n_{s}$. An instantaneous reheating $N_{re}\rightarrow 0$\
leads to the maximum temperature at the end of reheating $T_{re}$\ $\sim
10^{15}GeV.$

\subsection{Constraints on reheating temperature from gravitino abundance}

\label{sec:10}

The so-called \textquotedblleft gravitino problem\textquotedblright\ \cite%
{Gb15} is defined as when the gravitino decays forward to big bang
nucleosynthesis (BBN), its energetic daughters would destroy the light
nuclei via photo-dissociation, upsetting BBN's successful prediction. To
overcome this problem, an upper bound on the reheating temperature after
inflation is required. knowing that Thermal scatterings in plasma from the
reheating stage following inflation cause graviton formation \cite{Gb16}.
Thus it has been found that gravitino abundance is directly proportionate to
the reheating temperature $T_{re}$\ and can be approximated as \cite{Gb17}

\begin{equation}
Y_{\psi }\simeq 10^{-11}\left( \frac{T_{re}}{10^{10}GeV}\right) ,
\end{equation}%
where $Y_{\psi }$\ being the entropy density. This abundance should be low
so that the gravitino decay products do not destroy the light elements
successfully produced during BBN, thus we obtain the upper bound on the
reheating temperature such as \cite{Gb18} 

\begin{equation}
T_{re} \leq 10^{6-7}GeV.
\end{equation}

This result will help us understand more about the final reheating
temperature constraints since we previously concluded that reheating can be
considered as instantaneous if the temperature reached $T_{re}$\ $\sim
10^{15}GeV,$\ we can no longer consider this case in order to avoid the
gravitino abundance from reheating. Having this bound on $T_{re}$\ will
affect the preheating duration that we will compute in the next section,
from Eq.(\ref{Npre}) a lower value of reheating temperature can cause a
decrease in the $e$-folds number of preheating.

\subsection{Preheating case}

\label{sec:11}

As mentioned previously, the process of preheating happens in the early
stages of the Universe evolution. This is considered to be necessary since
as the universe expands, it cools down. Thus, right after inflation, there
must be a period to make it prepare thermally for next steps. By taking this
preheating period into account, we are able to find some constraints on the
duration $N_{pre}$ that we show to be independent of the (EoS) $\omega
^{\ast },$ this can be regarded as a possibility for gaining information
about the preheating period.

\begin{center}
\begin{figure}[H]
\centering
\includegraphics[width=15cm]{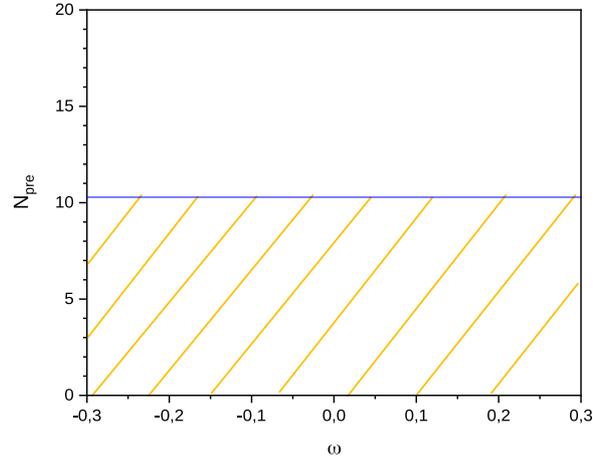}
\caption{Variation of $N_{pre}$ a function of equation of state $\protect%
\omega ^{\ast }$.}
\label{fig:8}
\end{figure}
\end{center}

In Fig. \ref{fig:8} we present a variation of $N_{pre}$ as functions of
(EoS) $\omega ^{\ast }$, the blue line is the maximum value that preheating
could take ($N_{pre}\approx 10$), bellow this value the region with yellow
lines, is estimated to be other possible preheating durations that one could
get if we consider $T_{re}<10^{7}$.

\begin{center}
\begin{figure}[H]
\centering
\includegraphics[width=15cm]{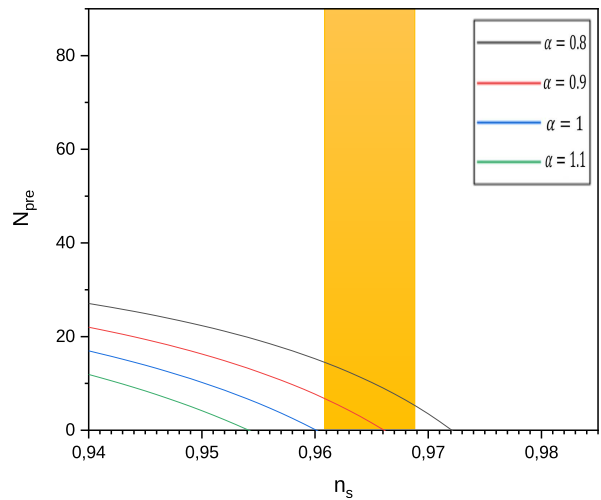}
\caption{Variation of $N_{pre}$ as a function of $n_{s}$ for differentes
values of parameter $\protect\alpha $ .}
\label{fig:9}
\end{figure}
\end{center}

In Fig. \ref{fig:9} we've computed $N_{pre}$ as a function of $n_{s} $, each
line represents a specific value of parameter $\alpha $, we chose $\omega
^{\ast }=0$ taking into consideration the result of the work \cite{Gb14},
that proved this value is necessary for an efficient preheating, knowing
that preheating duration is independent of the choice of (EoS) $\omega
^{\ast }$. The cases $\alpha =0.8$ and $\alpha =0.9$ are compatible with
observations. However, the values $\alpha =1$ and $\alpha =1.1$ \ doesn't
reproduce a preheating duration with compatible index spectral $n_{s} $
according to $Planck^{\prime }s$ data.

\section{Conclusion}

\label{sec:12}

In this work, we have studied the pre/reheating after supersymmetric
brane-world inflation. We reviewed reheating in the context of RSII
inflation by introducing the ratio $H_{0}/H$ and visualized the effect of
reheating temperature. We have also studied the preheating mechanism in the
hybrid brane inflation and set conditions on coupling constants for
efficient $\chi -$particle production. By calculating the reheating
temperature and pre/reheating durations quantified in terms of $e$-foldings
number, we show that preheating is related to the reheating temperature $%
T_{re}$\ \ that affect the parameter $N_{pre}$, and prove $N_{pre}$ to be
independente on the (EoS) $\omega ^{\ast }$. We've set constraints on\ the
parameters of D-term hybrid potential, for the same model we tested the
compatibility of the inflationary parameters $n_{s}$\ and $r$\ with Planck's
data. We proposed as well an investigation of the brane effect on the
reheating epoch. We have applied the form of the Hybrid D-term potentials
presented in previous works to study preheating and reheating constraints
and obtain information about pre/reheating from the Cosmic Microwave
Background. We have shown that the Hybrid D-term model show good
compatibility for $\alpha \leq 1$\ in the case of reheating while the
preheating case demand that $\alpha \leq 0.9$\ from recent observations. %
%
%

%
%

\end{document}